\documentclass[
    reprint,
    preprintnumbers,
    superscriptaddress,
    nofootinbib,
    amsmath,amssymb,
    aps,
    prl,
    floatfix,
    longbibliography]{revtex4-2}
    
    \usepackage{graphicx}
    \usepackage{subfigure}
    \usepackage{subcaption} 
    \usepackage{multirow}
    \usepackage[usenames,dvipsnames]{xcolor} 
    \usepackage{mathrsfs} 

    \usepackage{pgfplots}
    \usepackage[utf8]{inputenc}
    \usepackage{color}
    \usepackage{hyperref}
    \usepackage[normalem]{ulem} 
    \usepackage{physics}
    \usepackage{enumitem}
    \usepackage{comment}
    \usepackage{bm}
    \usepackage{aas_macros}
    \usepackage[capitalise]{cleveref}
    \usepackage{multirow}

\hypersetup{
  breaklinks = true,
  colorlinks   = true, 
  urlcolor     = Blue, 
  linkcolor    = Blue, 
  citecolor   = Blue 
}

\usepackage{float}

\newcommand{\oxford}{Astrophysics, University of Oxford, DWB, Keble Road, Oxford OX1 3RH, United Kingdom}

\newcommand{\qmulphysics}{Astronomy Unit, School of Physical and Chemical Sciences, Queen Mary University of London,
Mile End Road, London E1 4NS, United Kingdom}

\newcommand{\splitatcommas}[1]{%
  \begingroup
  \begingroup\lccode`~=`, \lowercase{\endgroup
    \edef~{\mathchar\the\mathcode`, \penalty0 \noexpand\hspace{0pt plus 1em}}%
  }\mathcode`,="8000 #1%
  \endgroup
}

\begin{document}

\title{Assessing cosmological evidence for non-minimal coupling}

\author{William J. Wolf}
\email{william.wolf@stx.ox.ac.uk}
\affiliation{\oxford}
\author{Carlos Garc\'ia-Garc\'ia}
\affiliation{\oxford}
\author{Theodore Anton}
\affiliation{\oxford}
\affiliation{\qmulphysics}
\author{Pedro G. Ferreira}
\affiliation{\oxford}

\begin{abstract}
The recent observational evidence of deviations from the $\Lambda$-Cold Dark Matter ($\Lambda$CDM) model points towards the presence of evolving dark energy. 
The simplest possibility consists of a cosmological scalar field $\varphi$, dubbed quintessence, driving the accelerated expansion.
We assess the evidence for the existence of such a scalar field. 
We find that, if the accelerated expansion is driven by quintessence, the data favour a potential energy $V(\varphi)$ that is concave, i.e., $m^2=d^2V/d\varphi^2<0$. Furthermore, and more significantly, the data strongly favour a scalar field that is non-minimally coupled to gravity (Bayes factor $\log(B) = 7.34 \pm 0.6$), leading to time variations in the gravitational constant on cosmological scales, and the existence of fifth forces on smaller scales. The fact that we do not observe such fifth forces implies that {\it either} new physics must come into play on non-cosmological scales {\it or} that quintessence is an unlikely explanation for the observed cosmic acceleration.
\end{abstract}

\maketitle


\textit{Introduction---}A new generation of cosmological surveys has allowed us to place much tighter constraints on the history of the Hubble rate. Until now, a model in which late time acceleration is driven by a cosmological constant, $\Lambda$, has been adequate to describe observations. However, new data are providing intriguing, but tentative, evidence for evolving dark energy  \cite{DESI:2024mwx, DESI:2025zgx}.

It is often useful to characterize dark energy in terms of its bulk properties. We can define an equation of state, $w\equiv P_{\rm DE}/\rho_{\rm DE}$, in terms of the pressure and energy density of the dark energy. $\Lambda$ has an equation of state $w=-1$, but evolving dark energy has an equation of state $w(a)$, which is a function of the scale factor of the Universe and is often approximated in terms of two parameters, $ w(a)\simeq w_0+w_a(1-a)$, known as the CPL parametrization \cite{Chevallier:2000qy, Linder:2002et}.
Current observations of the Cosmic Microwave Background (CMB), Baryon Acoustic Oscillations (BAO), and type Ia Supernovae (SNe Ia) seem to indicate that $w_a<0$  \cite{Planck:2018vyg, Aghanim:2019ame, ACT:2023dou, ACT:2023kun, Scolnic:2021amr, Rubin:2023ovl, DES:2024jxu, DESI:2024mwx, DESI:2025zgx}, i.e., that the equation of state is ``thawing'', or increasing, with time.

The simplest form of thawing dark energy -- quintessence as a minimally coupled scalar field -- is only marginally favoured over a cosmological constant and is not statistically favoured over parametric models of an evolving equation of state. In other words, while the data do seem to prefer a dark energy that evolves in time, the type of evolution being uncovered is \textit{not} well-described by standard quintessence \cite{Wolf:2024eph, Wolf:2024stt}.  
Thus, if one is to assume that a scalar field is the source of late time acceleration, one needs to look more broadly. 

We will show that current cosmological data favour quintessence which is non-minimally coupled to gravity through a term of the form $\xi\varphi^2 R$. While the presence of non-minimal coupling is not, in and of itself, striking, it does lead to some far-reaching consequences for physics on other scales. Hence, if the evidence for non-minimal coupling persists and strengthens, it will necessarily imply a reformulation of our understanding of dark energy in a wider physical context.

\textit{Scalar field dark energy---}There has been much focus on phenomenological parametrizations of dark energy \cite{Wolf:2025jlc, Giare:2024gpk, Lodha:2025qbg, DESI:2024kob, Payeur:2024dnq, Keeley:2025stf, Akthar:2024tua, Alho:2024qds, Shlivko:2025fgv}. While these approaches can be informative, they do not shed light on the microphysics of dark energy \cite{Wolf:2023uno}. In other words, they do not help in fighting the spectre of underdetermination \cite{Ferreira:2025fpn} and so will not lead us to the exact microphysical theory for dark energy. In this paper we want to go further and try to glean information about the microphysics from cosmological data. 

What we mean by ``microphysical theory'' is the action for the fundamental field that constitutes dark energy. A common assumption is that dark energy is in the form of a scalar field, which can generally be described by an action \cite{Park:2010cw}:
\begin{eqnarray}\label{eq:fullaction}
S=\int d^4 x\sqrt{-g}\left[\right.&\frac{M^2_{\rm Pl}}{2}F(\varphi)R-\frac{1}{2}G(\varphi)X 
-V(\varphi) 
\nonumber \\ 
&-J(\varphi)X^2+\left.{\cal L}_M\left(g_{\alpha\beta},\psi_M\right)\right],
\end{eqnarray}
where $g_{\alpha\beta}$ is the metric, $R$ the Ricci scalar, $\varphi$ the scalar field,   $X=\partial_\mu\varphi\partial^\mu\varphi$, and ${\cal L}_M$ is the action for matter.

Current data already constrain the terms in Eq.~\eqref{eq:fullaction}. Thawing quintessence ($F = G = 1$, $J = 0$) is not particularly favoured \cite{Wolf:2024eph, Wolf:2024stt}. Although it lies in the thawing regime ($w_a < 0$) and can yield a better $\chi^2$ fit than $\Lambda$CDM \cite{Berghaus:2024kra, Tada:2024znt, Shlivko:2024llw, Notari:2024rti, Borghetto:2025jrk, Berbig:2024aee, Luu:2025fgw, DESI:2025hce}, evidence-based comparisons are less optimistic. Depending on the SNe dataset, it is either disfavoured relative to $\Lambda$CDM or outperformed by standard $(w_0, w_a)$ models \cite{Wolf:2024eph, Payeur:2024dnq, Lodha:2025qbg, Wolf:2025jlc, Bhattacharya:2024hep, DESI:2025hce, Hossain:2025grx}. If dark energy evolves, the trend appears inconsistent with simple thawing quintessence models and instead suggests rapid evolution (suggested by the large, negative $w_a$ values favoured by the data), and possibly even phantom behavior in the past ($w < -1$) \cite{Berti:2025phi, DESI:2024aqx, DESI:2024kob, Lodha:2025qbg, Ye:2024ywg, Wolf:2024stt, Mukherjee:2024ryz, Ormondroyd:2025exu, Moghtaderi:2025cns, Ye:2025ulq}.

For the case of $F\neq 1$ \cite{Ye:2024ywg, Ye:2024zpk, Wolf:2024stt, Chudaykin:2024gol, Ferrari:2025egk, Chudaykin:2025gdn, Pan:2025psn, Wolf:2025jlc, Gannouji:2006jm} we can consider a general subset of models by expanding $F$ and $V$,
\begin{eqnarray}
    F(\varphi)&\simeq&1-\xi\frac{\varphi^2}{M^2_{\rm Pl}}, \nonumber \\
    V(\varphi)&\simeq&V_0+\beta\varphi+\frac{1}{2}m^2\varphi^2, \label{functions} 
\end{eqnarray}
with $G(\varphi)=1$ and $J(\varphi)=0$ (see appendix for more details on this expansion to quadratic order). Despite keeping ourselves ignorant about the specific quintessential action, we can at least try to answer two fundamental questions about the scalar field: ``what is the mass of the scalar field?'' and ``is it non-minimally coupled to Gravity''?

\begin{figure}[t]
   \centering
    {%
       \includegraphics[width=\columnwidth]{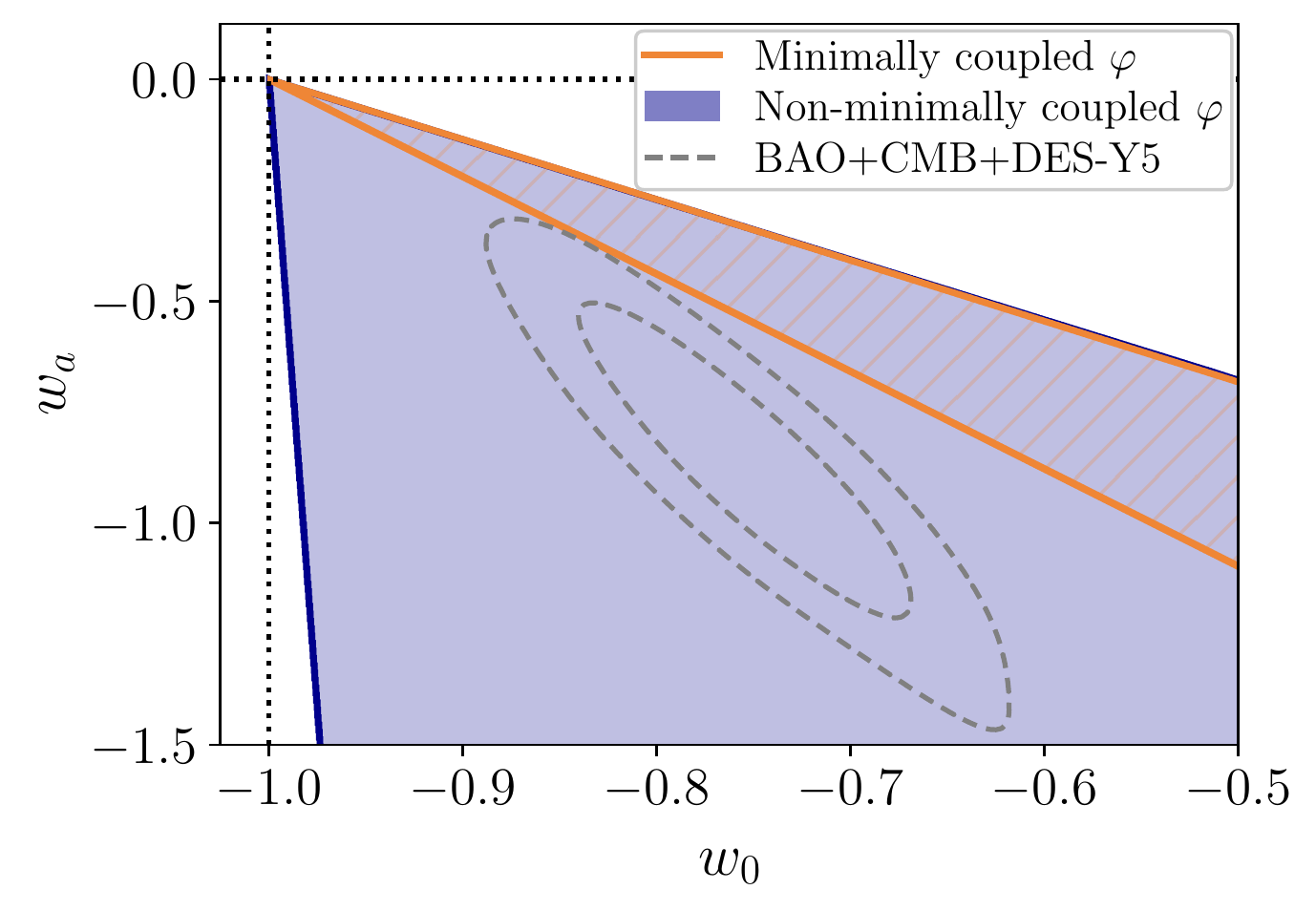}
    }
           \vskip -0.3in
    \caption{Projected theory priors into the $(w_0, w_a)$ plane for both minimally and non-minimally coupled dark energy by fitting $(w_0, w_a)$ to the theory prediction of the observables used in the data constraints.}\label{Fig:wedge}
    
\end{figure}

Dark energy described by Eqs.~\eqref{eq:fullaction} and \eqref{functions} typically thaws, as current data suggests, but behaves differently for $\xi=0$ versus $\xi\ne 0$. In the minimally coupled case, the field starts as potential-energy dominated, mimicking a cosmological constant, then evolves around $z\sim 0$–$2$ with $w_a<0$, though not steeply enough to fully match observations \cite{Wolf:2024eph, Wolf:2024stt}.
With non-minimal coupling ($\xi\neq0$), the field evolves like a minimally coupled field during radiation domination (given that $R\simeq 0$), yielding $w\simeq -1$. In the matter era, the effect of the non-minimal coupling is to push the field into the phantom regime ($w<-1$), then across the phantom divide as the field accelerates (see figure in appendix). We find that the combination of non-minimal coupling generating phantom behavior and $m^2=d^2V/d\varphi^2 < 0$ generating rapid thawing leads to more rapid variation in $w$ across the redshift range where most of the data is concentrated ($0.2 \lesssim z \lesssim 2$). This generates a more negative $w_a$, improving the agreement with data. One can see this in Fig.~\ref{Fig:wedge} where the minimal and non-minimally coupled theories have been projected onto the $(w_0, w_a)$ plane following the procedure in \cite{Wolf:2024eph, Wolf:2025jlc, Wolf:2024stt, Garcia-Garcia:2019cvr, Traykova:2021hbr}, which determines a theory prior in the $(w_0, w_a)$ parameters by directly using the errors associated with the observables.

\textit{Results---}We investigate the cosmological evidence for dynamical dark energy using a collection of data comprising DESI DR2 BAO \cite{Andrade:2025xyg, DESI:2025zpo}, Planck 2018 CMB temperature and polarization \cite{Planck:2018vyg, Aghanim:2019ame}, and ACT DR6 CMB lensing \cite{ACT:2023dou, ACT:2023kun}. These make our baseline data combination (BAO+CMB), which we supplement with SNe Ia data from Pantheon+ \cite{Scolnic:2021amr}, Union3 \cite{Rubin:2023ovl}, or DES-Y5 \cite{DES:2024jxu, DES:2025bxy}. For all data, we use the official likelihoods, as implemented in \texttt{Cobaya} \cite{Torrado:2020dgo}. For Planck, we use Planck PR3 \texttt{plik}. 
We use the nested sampler \texttt{polychord} \cite{Handley:2015fda} to derive the parameter posterior distributions and to calculate the Bayesian evidence
$   \log \mathcal{Z} = \log \int \mathcal{L}(D |\theta, M) P(\theta | M) \, d\theta,$
where $\mathcal{L}(D | \theta, M)$ is the likelihood,  $P(\theta| M)$ is the prior, and $\theta$ the sampled parameters. In order to provide an additional assessment of these models, we also compute the Deviance Information Criteria (DIC) defined as $\mathrm{DIC} = D(\bar{\theta}) + 2 p_D $, where $D(\theta) = -2 \log \mathcal{L}(\theta)$ and $p_D = \overline{D(\theta)} - D(\bar{\theta})$ is a term that penalizes additional complexity in the model. $\overline{D(\theta)}$ denotes the average deviance over the posterior and $D(\bar{\theta})$ denotes the deviance evaluated at the posterior mean.

For dark energy, we consider three models (see appendix for a discussion of parameter priors):
(i)~\textit{Non-minimally coupled quintessence}; 
(ii)~\textit{Minimally coupled quintessence}; corresponding to the model studied in \cite{Wolf:2024eph, Wolf:2023uno} where $V(\varphi)=V_0 + \frac{1}{2}m^2\varphi^2$; and
(iii)~the CPL parametrization.
We quantify their fit to the data by reporting the best fit $\chi^2 = -2 \log \mathcal{L}$ and its difference with respect to $\Lambda$CDM, $\Delta \chi^2_{X \Lambda} = \chi^2_X - \chi^2_\Lambda$. In order to account for the extra degrees of freedom, we use both the $\Delta \mathrm{DIC}_{X\Lambda}$ and the Bayes factor $\log B_{X \Lambda } = \log \mathcal{Z}_{X} - \log\mathcal{Z}_{\Lambda}$ to assess the supporting evidence in favour of a given model over $\Lambda$CDM. For both $\Delta \chi^2$ and $\Delta \mathrm{DIC}$, negative values indicate a preference for the model, whereas for $\log B$ positive values indicate a preference for the model. Note that on the Jeffrey's scale $\log B_{X \Lambda }> 5$ indicates strong evidence for model $X$ relative to $\Lambda$CDM \cite{Jeffreys1939, Trotta:2005ar}.
The results can be found in Table~\ref{Table: main_results}, where the $\log B_{X \Lambda }$ values all have an uncertainty of $\simeq 0.6$. 

\begin{table}[t]
    \centering
    \renewcommand{\arraystretch}{1.3}
    \begin{tabular}{@{}llccc@{}}
        \hline
        \hline
        & & Pantheon+ & Union3 & DESY5 \\
        \hline
        \multirow{3}{*}{\rotatebox[origin=c]{90}{$\log B$}}
        & Min. $\varphi$ & $-1.96$ & $+1.53$ & $+3.46 $ \\
        & $w_0 w_a$ & $-1.57 $ & $+4.02 $ & $+5.55 $ \\
        & Non-min. $\varphi$ & $+3.14 $ & $+6.45 $ & $+7.34 $ \\
        \hline
        \multirow{3}{*}{\rotatebox[origin=c]{90}{
        \begin{tabular}{c}
        $\Delta \chi^2$ \\
        ($\Delta \mathrm{DIC}$)
        \end{tabular}
        }}
        & Min. $\varphi$ & $-2.9$ ($-0.32$) & $-7.4$ ($-3.41$) & $-14.1$ ($-9.96$) \\
        & $w_0 w_a$ & $-8.0$ ($-4.89$)& $-14.0$ ($-12.2$) & $-18.2$ ($-16.4$)\\
        & Non-min. $\varphi$ & $-14.1$ ($-11.5$) & $-19.4$ ($-16.4$) & $-23.6$ ($-21.5$) \\
        \hline
    \end{tabular}
    \caption{Statistical comparison for dynamical dark energy models relative to $\Lambda$CDM from the combination of BAO, CMB, and different SNe samples. 
    }\label{Table: main_results}
\end{table}

For CPL, the $\Delta\chi^2_{w_0w_a \Lambda}$, $\Delta \mathrm{DIC}_{w_0w_a \Lambda}$, and  $\log B_{w_0w_a \Lambda}$ values are consistent with other works in the literature that have used similar datasets (e.g., \cite{Lodha:2025qbg, Giare:2024gpk, Wolf:2025jlc, DESI:2025hce, Wolf:2024eph}): in particular, $w_0w_a$ significantly improves the fit to all datasets, but is evidentially disfavoured with respect to $\Lambda$CDM when the Pantheon+ data are used, while it is evidentially favoured with Union3 and DES-Y5. For minimally coupled dark energy, while it can improve the fit to the data, it is the worst performing dynamical dark energy model. While it is evidentially disfavoured with respect to the Pantheon+ data (in agreement with \cite{Wolf:2024eph, DESI:2025hce}); we begin to see some moderate evidence for it emerge with the Union3 and DES-Y5 data. Given that these datasets significantly strengthen the preference for dynamical dark energy, we attribute this improvement in evidence to the fact that this dark energy model allows for more rapid evolution at low $z$ than standard minimally coupled quintessence models (when $m^2 < 0$) \cite{Wolf:2024eph, Wolf:2023uno, Shlivko:2024llw, Dutta:2008qn}.
Finally, the non-minimally coupled dark energy model provides the best fit to the data across the board. At first glance, this may not be surprising considering that it has more parameters than any of the other dark energy models considered, but the Bayesian evidence and DIC results indicate that the additional parameters justify their inclusion, with the model being strongly favoured by its Bayes factor and only receiving a modest DIC penalty. The non-minimally coupled model is the only model that is evidentially favoured over $\Lambda$CDM with respect to \textit{all} dataset combinations.

\begin{figure}[t]
   \centering
    {%
       \includegraphics[width=\columnwidth]{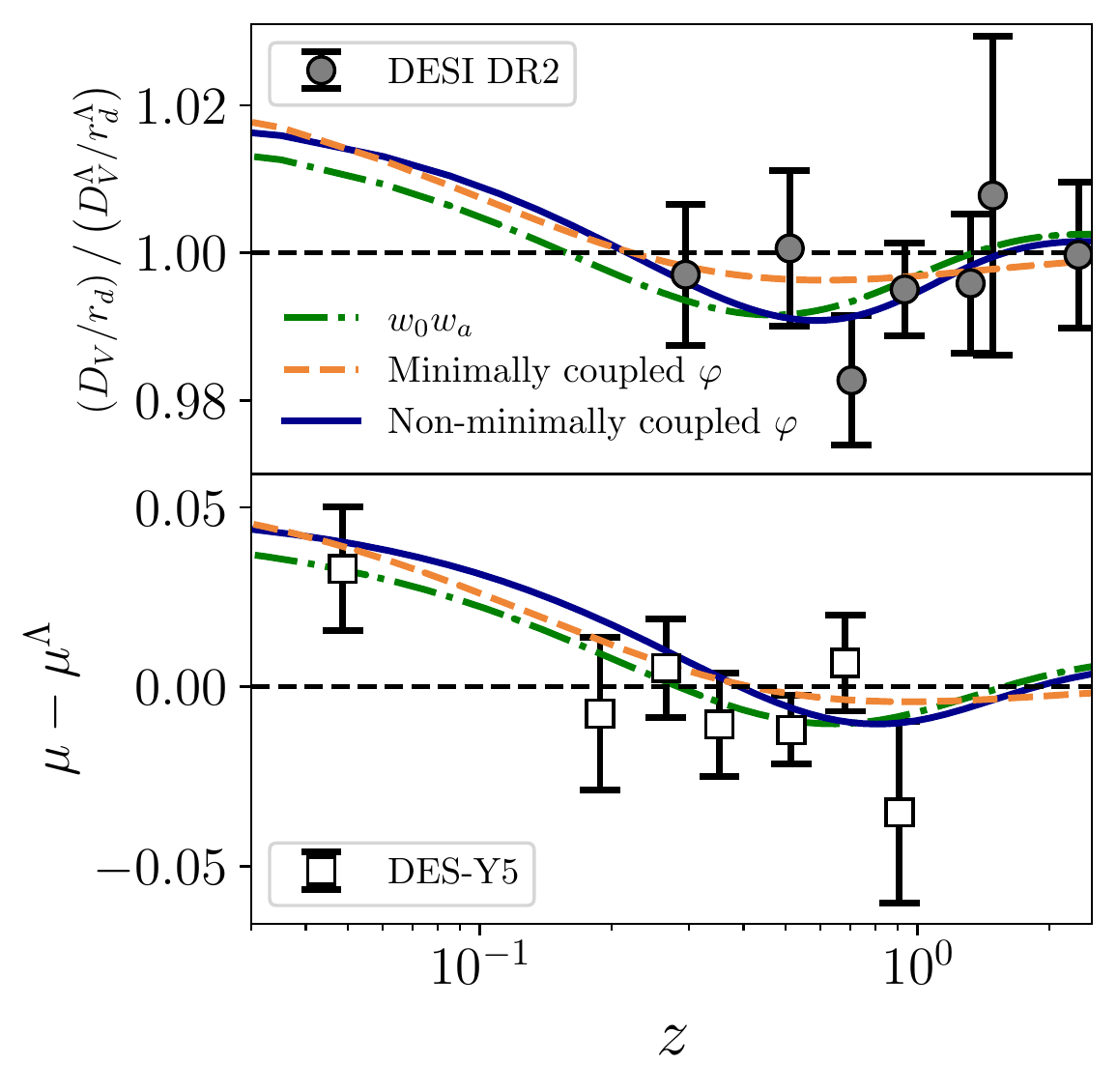}
    }
       \vskip -0.3in
    \caption{Observables from the best-fit dark energy models compared with the angle-averaged distance $D_{\rm V}$ and the distance modulus $\mu$, normalized to $\Lambda$CDM. 
    }\label{Fig:BF_v_data}
\end{figure}

Given that the BAO+CMB+DES-Y5 gives the strongest evidence for dynamical dark energy, we will dive a little further into this data combination. Fig.~\ref{Fig:BF_v_data} depicts both the DESI DR2 BAO data for the angle-averaged distance $D_V$ and the DES-Y5 SNe data for the distance modulus $\mu$ against the best fits for all dynamical dark energy models, all normalized to the best fit $\Lambda$CDM model. For visualization, the SNe data has been binned in redshift and we have calculated the
weighted average distance moduli and errors using the covariance matrix to include both statistical and systematic errors. See \cite[VII.B]{DESI:2025zgx} for a similar plot. Clearly, all dynamical models offer a significant improvement in fitting these distance measurements. In fact, they all perform similarly well in terms of their ability to fit the BAO and SNe data simultaneously, with $\Delta \chi^2_{X\Lambda} \simeq -15$ for every dynamical dark energy model. Where the CPL and non-minimally coupled dark energy separate themselves from minimally coupled dark energy is in their ability to simultaneously improve the fit to both the lower $z$ data and the CMB data over $\Lambda$CDM, which makes up the rest of the $\Delta \chi^2$ difference between these models and $\Lambda$CDM ($\Delta \chi^2_{X\Lambda} \simeq -3$ for CPL and $\simeq -7$ for the non-minimally coupled model).

\begin{figure}[t]
   \centering
    {%
       \includegraphics[width=\columnwidth]{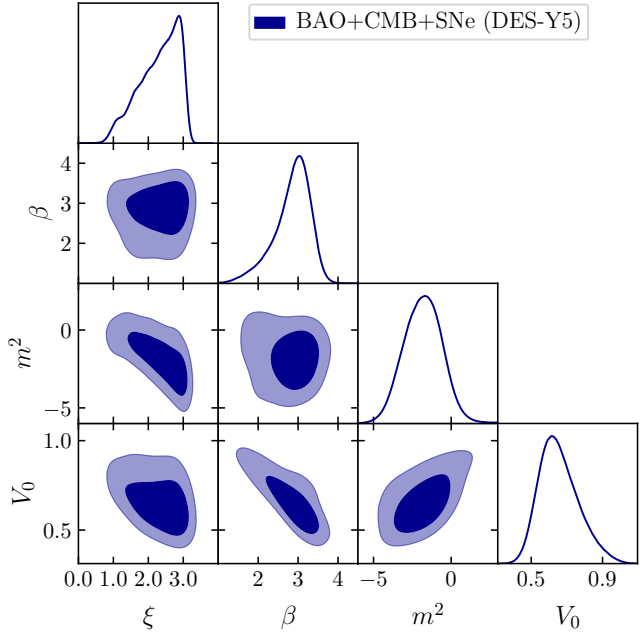}
    }
    \vskip -0.3in
    \caption{68\% and 95\% C.L. posterior distributions for the non-minimally coupled dark energy parameters.}\label{Fig:Posteriors}
\end{figure}

\begin{figure*}[ht]
    \centering
    \includegraphics[width=\textwidth]{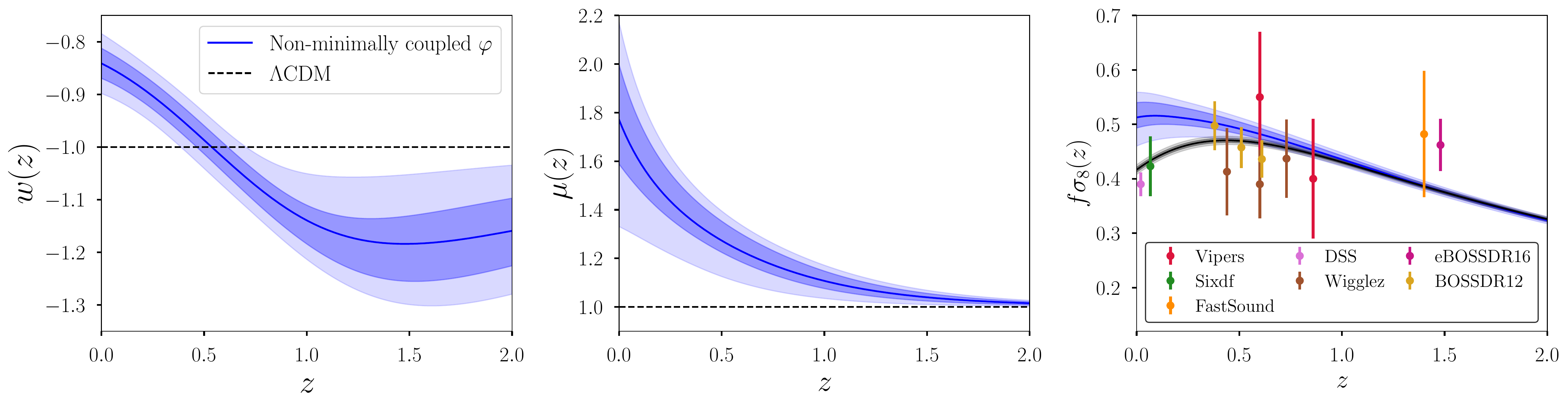}
    \caption{Reconstruction of the equation of state $w(z)$, $\mu(z)$, and $f\sigma_8(z)$ obtained in Fig.~\ref{Fig:Posteriors}. The $f\sigma_8$ measurements were not used in obtaining the constraints and are plotted for reference. One can see that this non-minimally coupled model is reasonably compatible with these $f\sigma_8$ data points \cite{2012MNRAS.423.3430B, Okumura:2015lvp, 2012MNRAS.425..405B, BOSS:2016wmc, Stahl:2021mat, Mohammad:2017lzz, eBOSS:2020yzd}.}
    \label{Fig:reconstructed}
\end{figure*}

Fig.~\ref{Fig:Posteriors} shows the constraints of the non-minimally coupled scalar field parameters. We find 
$\xi = 2.31^{+0.75}_{-0.34}$,
$\beta = 2.85^{+0.53}_{-0.29}$,
$m^2 = -1.8 \pm 1.2$, and
$V_0 = 0.654^{+0.086}_{-0.13}$.
Note that both minimally and non-minimally coupled dark energy models strongly favour $m^2=d^2V/d\varphi^2<0$, as concave potentials allow for more rapid variation in the scalar field \cite{Wolf:2023uno, Wolf:2024eph, Shlivko:2024llw, Dutta:2008qn}. Such potentials are unbounded from below, absent of other corrections; however, this is not relevant for cosmological constraints as we find that the field excursion is  $\Delta \varphi/M_{\rm Pl} \simeq {\cal O}(0.1)$ and never enters a pathological regime.
Finally, the left hand panel of Fig.~\ref{Fig:reconstructed} shows the constraints on $w(z)$. As anticipated, the posterior distribution of $w(z)$ is consistently below $-1$ by more than $2\sigma$ in the matter dominated era, crossing the phantom divide at about $z\sim 0.5$ and reaching $w(z=0) =-0.841^{+0.029}_{ -0.028}$ today, $5.7\sigma$ away from $w=-1$. This quick transition over this range of redshifts is what allows this model to significantly improve the fit to both the early (CMB) and late time (BAO and SNe) data simultaneously, and resembles the evolution favoured by the parametric dark energy models (e.g., \cite[Fig.~4]{Lodha:2025qbg}). Furthermore, we also see in the central panel that the non-minimal coupling has a significant impact on the strength of gravity, parametrized by $\mu(z)$, with $\mu(z=0) =  1.77^{+0.23}_{-0.18}$, $4.3\sigma$ away from the General Relativity (GR) with $\mu = 1$. These cosmological results have profound implications for the nature of gravity.

\textit{Ancillary Gravitational Consequences---}The non-minimal coupling of the scalar field to the metric brings with it a host of undesirable consequences and it would be remiss of us not to consider them. On the largest scales, we have that the effective gravitational constant, $G_{\rm eff}$ (as opposed to the bare gravitational constant which appears in the action, $8\pi G_0=M^{-2}_{\rm Pl}$) is time varying, ${\dot G_{\rm eff}}/G_{\rm eff}=2\xi\varphi{\dot \varphi}/(1-\xi\varphi^2)$. We find that, for parameters that can best fit the cosmological data, ${\dot G_{\rm eff}}/G_{\rm eff}\simeq 10^{-11}/{\rm yr}$.  Constraints on this quantity (which are independent of any assumptions about the cosmological model) should not be confused with the constraints that come from the Solar System or compact binaries on the Newton-Poisson $G_{\rm N}$ 
\cite{Wolf:2019hun, Lagos:2020mzy, Dalang:2019fma} (although they are around the same order of magnitude \cite{Uzan:2024ded}).

As we go further down in scale, we have that the non-minimal coupling will affect the gravitational constant which enters in the Newtonian limit, in the Newton-Poisson equation. We then have that $\nabla^2 \Phi=4\pi G_0 \mu(z) {\bar \rho}_{\rm M}\delta_{\rm M}$ where $\Phi$ is the Newtonian potential, ${\bar \rho}_{\rm M}$ and $\delta_{\rm M}$ are the background and fractional density perturbation of matter and $\mu(z)$ captures the time evolution of Newton's constant on the scales of gravitational clustering. We can see the impact of this time variation in two distinct ways. The central panel of Fig.~\ref{Fig:reconstructed}  shows the inferred $\mu(z)$ from CMB+DESI+DESY5 data. For $z<0.6$, $\mu(z)$ deviates from the GR value by over two standard deviations ($4.3\sigma$ today), suggesting a preference for modifications to gravitational clustering in the Newtonian regime (see also \cite{Pogosian:2021mcs} for a compatible model-agnostic reconstruction). The right hand panel shows $f\sigma_8(z)$, where $f=d\ln \delta/d\ln\eta$ is the growth rate and $\sigma_8$ is the mass variance on scales of $8h^{-1}$ Mpc. The non-minimally coupled model is  consistent with higher redshift measurements, although less so for the lowest redshift measurements. Note, however, that current data on structure growth are still too imprecise to tightly constrain departures from $\Lambda$CDM; furthermore, an accurate covariance matrix for how these measurements correlate with distance measurements is still lacking, hindering any attempts at a combined analysis.

Non-minimally coupled scalar fields are tightly constrained on small scales, with no evidence of fifth forces in astrophysical systems \cite{adelberger2003tests, Will:2014kxa}. Specifically, PPN parameters are strongly bounded: Cassini data \cite{bertotti2003test} constrain the light-bending parameter to $\gamma_{\rm PPN} - 1 = (2.1 \pm 2.3)\times 10^{-5}$, and MESSENGER \cite{park2017precession, 2012Sci...336..214S} bounds the nonlinearity parameter to $\beta_{\rm PPN} - 1 = (-2.7 \pm 3.9)\times 10^{-5}$. These imply $\xi(\varphi_0/M_{\rm Pl})^2 \le 6 \times 10^{-6}$ and $\xi \le 1$, in tension with the cosmological requirement $\xi(\varphi_0/M_{\rm Pl})^2  \sim 0.1$. In the Einstein frame, matter couples via ${\tilde g}_{\alpha\beta} \sim [1 - \xi(\psi/M_{\rm Pl})^2]g_{\alpha\beta}$, leading to time- and space-varying constants that are observationally constrained.

Fifth forces can be suppressed on small scales, as seen from the action in Eq.~\eqref{eq:fullaction}. Adding the EFT-allowed term $\frac1{\Lambda^4}\int d^4 x\sqrt{-g} X^2$ enables gravitational screening that protects small scales from large deviations \cite{Khoury:2003rn, Brax:2021wcv, Sakstein:2017pqi, Vardanyan:2023jkm, Ye:2024zpk, Briddon:2024ftz}. This leads to Vainshtein screening \cite{Vainshtein:1972sx}, distinct from Galileon models, due to the explicit non-minimal coupling to the Einstein-Hilbert term. The fifth force becomes sensitive to the local scalar field and, in the screened regime, takes the form $F_5/F_{\rm N} \propto (r/r_{\rm V})^{4/3}$ \cite{quintessence_nonminimal_2025_inprep}, where $r_V = [9M/(16\pi \Lambda^2 M_{\rm Pl})]^{1/2}$, and $F_{\rm N} \propto r^{-2}$. For comparison, cubic Galileons yield $F_5/F_{\rm N} \propto (r/r_{\rm V})^{3/2}$ \cite{Kobayashi:2019hrl}. With $\Lambda \lesssim 10^{-2}\,{\rm eV}$, Solar System screening ($r_{\rm V} \gtrsim 10^{-3}\,{\rm pc}$ for $M \sim M_{\odot}$) is achievable without altering cosmic expansion. Another alternative is to break the universality of the non-minimal coupling and assume that the dark energy, in the Einstein frame, is {\it only} coupled to the dark matter. Such a scenario corresponds to an interacting dark matter/dark energy model \cite{Gomez-Valent:2020mqn, Amendola:1999er, Giare:2024smz, Chakraborty:2025syu, Shah:2025ayl, Silva:2025hxw, Khoury:2025txd} and might avoid some of the constraints on fifth forces. However, as with screening, it would be another (more contrived) version of non-minimal coupling.

The only way to avoid non-minimal coupling may be through a radical reformulation of the kinetic structure of the scalar field. A purely shift-symmetric theory, replacing the kinetic term by a more complicated function, $X\rightarrow K(X, \Box\varphi)$, leads to  $w<-1$ throughout and the resulting $(w_0, w_a)$ does not overlap with the current constraints \cite{Traykova:2021hbr}. Breaking the shift symmetry by, for example, adding a potential, may make such theories more consistent with the data, but will in turn lead to other problems, such as superluminal motion and instabilities. We will explore these models in future work.


\textit{Discussion---}In this letter we have shown that {\it if we assume that the accelerated expansion is driven by a dynamical scalar field}, there is compelling statistical evidence (by any measure) that it must be non-minimally coupled to gravity and that its potential $V$ satisfy $d^2V/d\varphi^2<0$. The non-minimal coupling will drive time variation of the effective gravitational constant, affect the growth rate of structure, and lead to fifth forces on much smaller scales. Such forces are tightly constrained through a range of different methods on laboratory and Solar System scales. Given the {\it absence} of any evidence for fifth forces, this means that additional new physics must be invoked to screen the presence of a non-minimal coupling. 

Such a non-minimal coupling will naturally arise in the effective action of a theory coupled to gravity \cite{Kaiser:2015usz, Buchbinder1992, Birrell_Davies_1982, Esposito-Farese:2000pbo, Park:2010cw, Callan:1970ze, Markkanen:2013nwa, Nojiri:2010wj}. Furthermore, theories with extra dimensions may lead to non-minimal couplings, with one caveat: given the substantial deviation of $w(a)$ from $-1$, we inevitably have substantial field excursions. In this case we find that $\Delta\varphi/M_{\rm Pl} \sim {\cal O}(0.1)$ which is manageable, yet far too large to be accommodated in some of the better motivated theories of dark energy. Furthermore, the screening scale $\Lambda$ will be substantially larger than $(H_0 M_{\rm Pl})^{1/2}$ and thus may lead to strong coupling problems.

The quintessence hypothesis clearly leads us down a path of ever increasing complexity, and the consequences are sufficiently jarring, that it behooves us to step back and reconsider our assumptions. A key assumption is that the dark energy is in the form of a scalar field. One might want to consider the handful of proposals which do not reduce to a scalar field on large scales (or an effective description in terms of a scalar-tensor theory) \cite{Clifton:2011jh}. Another possibility is that the details of the accelerated expansion we are inferring is a result of our mismodelling the cosmological space-time and, as a result, 
observables may be non-trivially affected by the inhomogeneous nature of the cosmos  \cite{Buchert_2012, Buchert_2015, Rasanen_2006, Heinesen_2021, Seifert_2024}. We should also revisit the assumptions that have gone into the analysis of the data -- the presence or absence of a cosmic dipole, the role of the low redshift supernovae, the discrepancy between different supernovae and between different BAO samples, etc \cite{Efstathiou_2024, Colgain:2024mtg, Carloni:2024zpl, Luongo:2024zhc, Tang:2024lmo, DES:2025tir, Ghosh:2024kyd, Notari:2024zmi, Wang:2024dka, Wang:2024rjd, Dinda:2024kjf, Chan-GyungPark:2024mlx, Chan-GyungPark:2024brx, Giare:2025pzu, Banik:2025dlo, Bansal:2025ipo, Dinda:2024ktd, Cortes:2024lgw, Gao:2025ozb, Peng:2025nez, Mukherjee:2025fkf, Jiang:2024xnu, Huang:2025som, Gialamas:2024lyw, Dinda:2025iaq, Secrest:2020has, Secrest:2022uvx, Sah:2024csa}. We will watch with interest as the various analyses are independently scrutinized, and await the expected increase in the quality and quantity of the data from the new generation of surveys.

\textit{Acknowledgements---}We thank D. Alonso, J. Conlon, E. Copeland, H. Desmond, D. Kaiser, K. Lodha, J. Noller, M. Vincenzi and G. Ye for many valuable conversations. WJW is supported by St.~Cross College, University of Oxford. CGG is supported by the Beecroft Trust.
TA acknowledges support from a Balzan Fellowship. TA is grateful to the Astrophysics Department at Oxford and New College Oxford for their hospitality throughout his visit.
PGF is supported by STFC and the Beecroft Trust. For the purposes of open access, the authors have applied a Creative Commons Attribution (CC BY) license to any Author Accepted Manuscript version arising.

\newpage
\section{Appendix}
\textit{Units and initial conditions---}For the parameters in the non-minimally coupled dark energy model given by Eqs.~\eqref{eq:fullaction} and \eqref{functions}, $\xi$ is dimensionless, $V_0$ is in units of $M_{\rm Pl}^2 M^2_{\mathrm H}$ (where $M_{\mathrm H} = H_0/h$), $\beta$ is in units of $M_{\rm Pl} M^2_{\mathrm H}$, and $m^2$ is in units of $M^2_{\mathrm H}$. We use \texttt{hi\_class} \cite{hi_class1,hi_class2,CLASS} to compute the theory predictions, starting the field at $\varphi_{\rm ini}=0$ and ${\varphi'}_{\rm ini} = 10^{-3}$ to ensure it starts rolling. Current data only probe a small part of the potential $V(\varphi)$ and are not sensitive to its full shape. In fact, it is the shape of the potential at $|\varphi| > |\varphi_{\rm ini}|$ 
that determines the field dynamics. As a consequence, there is little loss of generality by setting $\varphi_{\rm ini} = 0$. Finally, we tune $V_0$ to ensure that the Friedmann equations close (i.e., ensure that the sum of all fractional densities $\sum_i \Omega_i = 1$).

\textit{Quadratic approximation---}Although restricting $F$ and $V$ to quadratic functions may seem limiting, observables are sensitive to only a narrow range of the scalar field's evolution ($\Delta\varphi/M_{\rm Pl} \sim {\cal O}(0.1)$), which makes the expansion in Eq.~\eqref{functions} sufficient. In order to prove this, we expanded the potential out to the cubic term to confirm that this term does not significantly affect the evolution of the model. We found that including a cubic term with a coefficient of the same order as the quadratic term only imparts percent level deviations on the model's equation of state even today $(a=1)$, and furthermore confirmed that including this term does not change the model's fit to the data ($\Delta \chi^2 \simeq 0.15$) or the mean values of the posterior constraints on the leading order model parameters. Moreover, the data does not have constraining power over this cubic term, recovering its prior range, centered at zero. Finally, note that we used the freedom to shift the scalar field to remove the linear term in the non-minimal coupling, as this merely corresponds to a constant shift in the scalar field.  

This approach captures a broad class of non-minimally coupled scalar field models, serving as a model-agnostic parametrization valid for any thawing scalar field with a non-derivative non-minimal coupling. For instance, it includes models with exponential and hilltop potentials \cite{Ye:2024ywg, Ye:2024zpk, Wolf:2024stt}, as they admit an expansion like Eq.~\eqref{functions}.

\begin{figure}[t]
   \centering
    {%
       \includegraphics[width=\columnwidth]{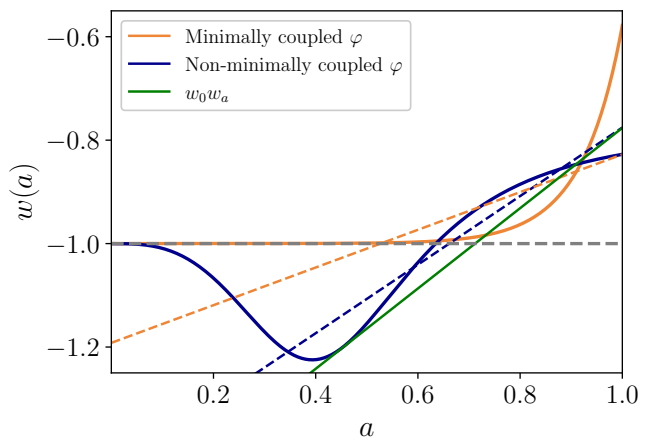}
    }
    \vskip -0.3in
    \caption{Best fit models for the various dark energy theories considered in this letter (solid lines), alongside their best-fit $(w_0,w_a)$ parameters (dashed lines) determined by fitting the CPL model to their observables predictions in the manner described in \cite{Wolf:2024eph, Wolf:2025jlc}.  }\label{Fig:EOS}
\end{figure}

\textit{Non-minimal dynamics---}Fig.~\ref{Fig:EOS} depicts the evolution of the best fit models for all dark energy models considered in this letter. Note also that the best fit non-minimally coupled model closely tracks the best fit CPL model, as this model can undergo significant, rapid evolution in the regime where most of the data is ($z\in[2.0,0.2]$), whereas the minimally coupled quintessence model has most of its dynamical evolution in very recent epochs, largely outside the regime where the data has most of its constraining power.

Note also, in this figure one can see that the non-minimally coupled model initially behaves approximately as a minimally coupled scalar field during radiation domination ($R\simeq0$) with $w\simeq-1$, then subsequent evolution drives the effective equation of state into the phantom regime, and finally the potential takes over to cause the field to thaw. To see this more explicitly, one can also schematically write the equation of state for this model in the following way \cite{Madsen:1988ph},
    \begin{equation}
        w = \frac{P}{\rho} = \frac{\frac{1}{2} \dot\varphi^2 - V + \xi F(\varphi\dot \varphi, \dot\varphi^2, \varphi\ddot\varphi)}{\frac{1}{2} \dot\varphi^2 + V + \xi G(\varphi\dot \varphi, \dot\varphi^2, \varphi\ddot\varphi)},
    \end{equation}
where $F$ and $G$ are functions of the scalar field and its time derivatives. At early times, the Hubble friction freezes the field ($\dot{\varphi} \simeq \ddot{\varphi} \simeq 0$) and the equation of state approximates that of a minimally coupled model, and then once the field begins rolling the equation of state can undergo the evolution we have discussed.

\textit{Sampling and priors---}
For dark energy, we consider three models:
(i)~\textit{Non-minimally coupled quintessence}, with 
\(\xi \in [0, 4.0]\), 
\(\beta \in [0, 10]\), and 
\(m^2 \in [-10, 10]\);
(ii)~\textit{Minimally coupled quintessence}, corresponding to the model studied in \cite{Wolf:2024eph, Wolf:2023uno} where $V(\varphi)=V_0 + \frac{1}{2}m^2\varphi^2$, with 
\(V_0 \in [-0.5, 1.5]\) and 
\(m^2 \in [-100, 5]\); and
(iii)~the CPL parameterization, with 
\(w_0 \in [-2, 0]\) and 
\(w_a \in [-2.5, 1.5]\). 
We adopt uniform priors on all cosmological and dark energy parameters, varying the standard cosmological parameters 
$\{\Omega_b h^2, \Omega_m, H_0, \ln 10^{10} A_s, n_s, \tau\}$ while keeping the neutrino mass fixed to \(\sum m_\nu = 0.06\,\mathrm{eV}\). 

Bayesian evidence is of course sensitive to the choice of priors. For the non-minimally coupled model, we choose wide priors over a range that keeps our model coefficients at $\mathcal{O}(1)$ (which again we think of as Taylor expansion coefficients). As noted in \cite{Lodha:2025qbg}, offsetting a Bayes factor of $\Delta \log \mathcal{Z} \simeq 5$ in the case of CPL would require a volume prior expansion of $\simeq 100$. Here, offsetting a Bayes factor of $ \Delta \log \mathcal{Z} \simeq 7$ would require expanding the prior volume by $\simeq 1000$. Similarly, the prior volume could be expanded by a factor of $\simeq 10$ and the evidence for the non-minimally coupled dark energy model would remain strong.
The case for minimally coupled thawing quintessence is slightly different. As the likelihood is widely insensitive to the negative end of $m^2$ (see \cite[Fig.~6]{Wolf:2024eph}), and the uniform prior on $m^2$ is normalized, the Bayesian evidence remains approximately constant when modifying the lower limit of the prior range  for $m^2$. 

\bibliography{refs}

\end{document}